\documentclass{SCIS2024}

\begin{document}
\ArticleType{PERSPECTIVE}
\Year{2024}
\Month{}
\Vol{}
\No{}
\DOI{}
\ArtNo{}
\ReceiveDate{}
\ReviseDate{}
\AcceptDate{}
\OnlineDate{}

\title{Next Generation Multiple Access for    IMT Towards  2030 and Beyond}{Title for citation}

\author[1]{Zhiguo Ding}{{zhiguo.ding@manchester.ac.uk}}
\author[2]{Robert Schober}{}
\author[3]{Pingzhi Fan}{}
\author[4]{H. Vincent Poor}{}

\AuthorMark{Author A}



\address[1]{Dept. CCE, Khalifa University, Abu Dhabi, UAE, and   Dept. EEE,   University of Manchester,  Manchester, U.K}
\address[2]{  Institute for Digital Communications, Friedrich Alexander-University Erlangen-Nurnberg (FAU), Germany}
\address[3]{    CSNMT Int. Coop. Res.
Centre (MoST), Southwest Jiaotong University, Chengdu 610032, China}
\address[4]{Department of Electrical and Computer Engineering, Princeton University,
Princeton, NJ 08544, USA}

\maketitle
\vspace{-2em}
\begin{multicols}{2}
\noindent

Multiple access techniques are fundamental to the design of wireless communication systems, since many crucial components of   such systems depend on the   choice of the multiple access technique [1]. For example, the use of orthogonal frequency-division multiple access (OFDMA)   simplifies  the physical layer design, where complicated channel estimation and equalization, which are   mandatory for a non-OFDMA   system   for combating    frequency selective fading,  are no longer needed. The use of OFDMA has also revolutionized   the design of the upper layers of a communication system, e.g.,  sophisticated scheduling and  resource allocation schemes  have  been introduced   by exploiting  the features of    OFDMA. 

Because of the     importance of multiple access, there has been an ongoing quest during the past decade to develop    next generation multiple access (NGMA).   Among those potential candidates for  NGMA, non-orthogonal multiple access (NOMA) has received significant attention from both the industrial and academic research communities,  and   has been   highlighted in   the recently published International Mobile Telecommunications (IMT)-2030 Framework as follows:  ``for multiple access, technologies including non-orthogonal multiple access (NOMA) and grant-free multiple access are expected to be considered to meet future requirements" [2]. In particular, the  literature of NOMA has demonstrated  its great potential   to  support the key usage scenarios of IMT-2030, namely  massive communication, ubiquitous connectivity, integrated sensing and communication (ISAC), and hyper reliable and low-latency communication  [3]. However, there is still no consensus in the research community about how exactly    NOMA assisted NGMA should  be designed. The aim of   this perspective is to illustrate the   three key features that NOMA assisted   NGMA should have,   as detailed in the following.

\begin{figure*}[t]
\centering
\includegraphics[width= 1\textwidth]{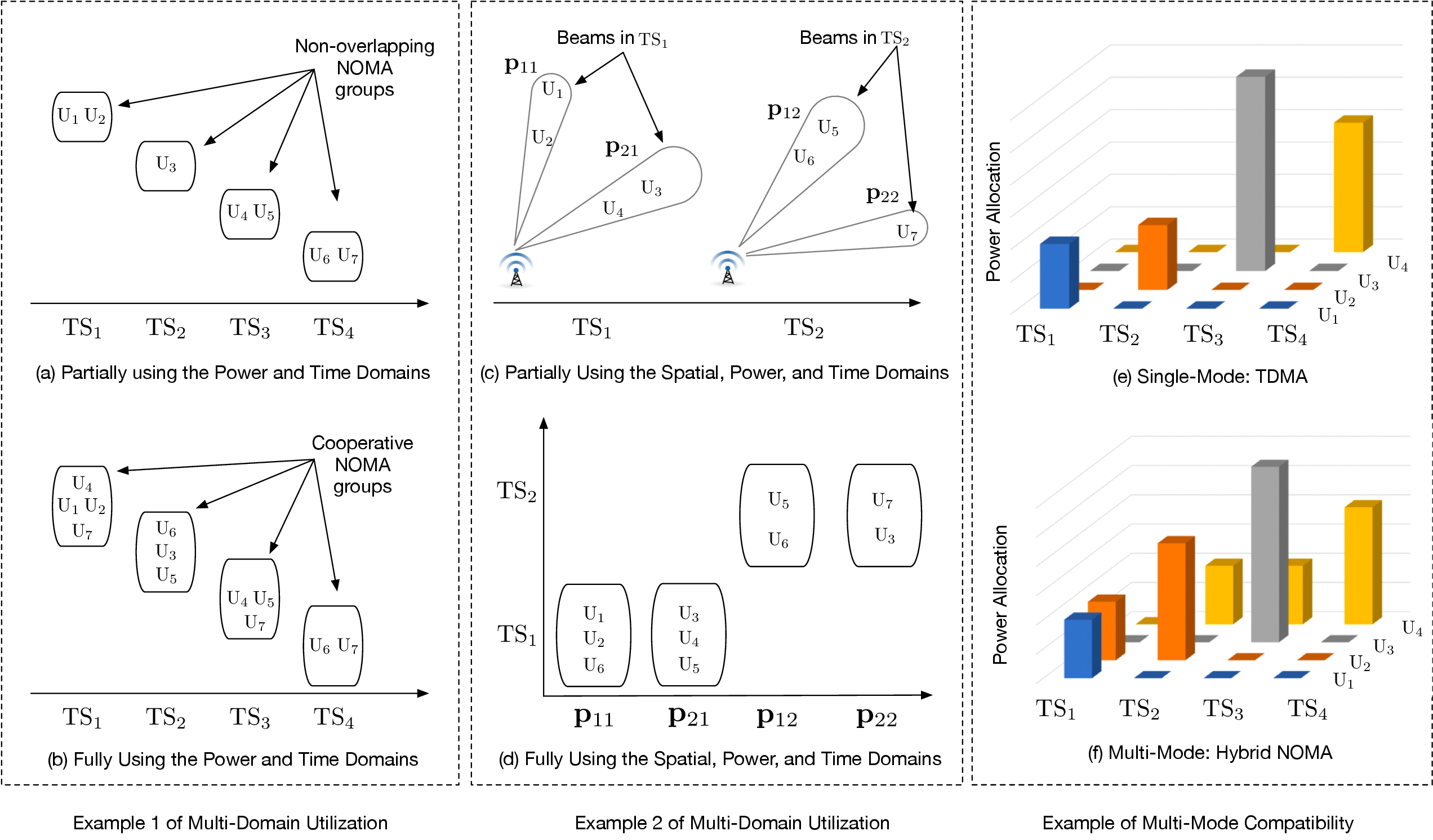}
\caption{Illustrations of the features of multi-domain utilization and multi-mode compatibility.}
\label{fig1}\vspace{-1em}
\end{figure*}

 
 {\it Multi-Domain Utilization:} Conventional NOMA and orthogonal multiple access (OMA)  have been designed by focusing on the efficient use of the bandwidth resource blocks available in a single domain, e.g., time division multiple access (TDMA) relies on      orthogonal resource blocks in the time domain, i.e., time slots, and many forms of   NOMA are designed to utilize the power domain for multiple access.  There have been some new forms of NOMA that  can realize partial multi-domain utilization, as illustrated  in Figs. 1(a) and 1(c). For example, Fig. 1(a) shows a type of NOMA that  utilizes both the power and time domains,   by first dividing users into multiple non-overlapping NOMA groups and then serving different groups in different time slots. One drawback of this type of NOMA is  that each user has access to a single time slot only, and hence the overall system performance can be improved with full multi-domain utilization, i.e., each user is provided   access to multiple time slots as shown in Fig. 1(b). The concept of multi-domain uitlization can be generalized by including  the spatial domain, as shown in Figs. 1(c) and 1(d).  In particular,   a typical combination of NOMA, TDMA, and space division multiple access (SDMA) is shown in Fig. 1(c). The resources from the space, power, and time domains can be  used in a more flexible and efficient manner  via  full multi-domain utilization  as shown in Fig. 1(d), where each user is not constrained to    a single time slot.

 {\it Multi-Mode Compatibility:}  Ideally,   NGMA   should be compatible with  different existing multiple access techniques, such that a dynamic coexistence between different multiple access modes can be supported. One way to support this multi-mode compatibility is to implement NOMA as an add-on to OMA, which leads to the concept of hybrid NOMA [4]. The key idea of hybrid NOMA is  illustrated in  Figs. 1(e) and 1(f). In particular, Fig. 1(e) shows an example of TDMA, where  four users are served  in four time slots  individually. Fig. 1(f) shows  a hybrid NOMA scheme that  enables the  dynamic implementation of OMA and NOMA. In particular, both users $1$ and $3$ adopt   OMA   and use a single time slot each, whereas users $2$ and $4$ adopt  NOMA.  In other words, hybrid NOMA can realize a harmonic and efficient integration of different multiple access techniques, where a user can freely choose a multiple access technique  based on its own capability and quality of service requirements.   We   note that multi-mode compatibility also has  the benefit that it can   be deployed in   existing OMA based networks, without the requirement of changing  the current standards.  
 
 

 {\it Multi-dimensional Optimality:} The aforementioned two features   lead to this new feature from the optimization perspective. For example,   single-dimensional  optimization is used to  carry  out   resource allocation for the  conventional designs shown in Figs. 1(a) and 1(c), i.e., the system optimization is carried out with the constraint that each user   has access to a single time slot only.  On the other hand, the multi-domain utilization schemes shown in Figs. 1(b) and 1(c) rely on     multi-dimensional    optimization, e.g., the system optimization is carried out without the constraint that each user can  use only a single time slot. From the optimization perspective, an optimization problem with fewer  constraints yields better performance, which is the reason why the multi-dimensional optimization schemes are expected to yield better performance than   single-dimensional ones.  We note that   multi-dimensional optimization problems are more challenging to solve than single-dimensional optimization problems, as illustrated with the following example.  
For   resource allocation   carried out within a single dimension, i.e., a single time slot,  a simple example of NOMA resource allocation   is provided   as follows:
  \begin{problem}\label{pb:1} 
  \begin{alignat}{2}
\underset{ P_i\geq 0 }{\rm{min}} &\quad      f_0(P_1, P_2)  \nonumber
\\ s.t. &\quad \log\left(1+ \frac{P_1}{P_2+1}\right)   \geq  R,  \label{1tst:2}  
  \end{alignat}
\end{problem}  
\noindent where $f_0(P_1, P_2) $ is an arbitrary resource allocation objective function, $R$ is the target data rate,   $P_i$ is user $i$'s transmit power, and $\log\left(1+ \frac{P_1}{P_2+1}\right) $ is user $1$'s data rate if user $2$'s signal is treated as noise. 
Problem \ref{pb:1} can be straightforwardly  solved since the constraint $\log\left(1+ \frac{P_1}{P_2+1}\right)   \geq  R$ can be converted to an affine form: $  P_1    \geq  (P_2+1)(e^R-1)$. 
For multi-dimensional optimization, we assume instead  that  user $1$ can   use one additional time slot with the transmit power denoted by $P_0$. The corresponding optimization problem can be formulated   as follows:
  \begin{problem}\label{pb:2} 
  \begin{alignat}{2}
\underset{ P_i\geq 0 }{\rm{min}} &\quad      f_0(P_1, P_2, P_0) \nonumber
\\ s.t. &\quad \log\left(1+ \frac{P_1}{P_2+1}\right) +\log(1+P_0)   \geq  R.  \label{2tst:2}  
  \end{alignat}
\end{problem}  
\noindent Comparing  Problems \ref{2tst:2}  and   \ref{1tst:2}, we observe that the difference between the two optimization problems seems trivial, where the simple term $\log(1+P_0)$ is added to the constraint of Problem  \ref{pb:2}. However, solving Problem \ref{pb:2} is significantly more challenging than solving  Problem \ref{pb:1}, since it is not a trivial task to convert constraint  \eqref{2tst:2}  to an affine or convex form.

{\it Open Problems for Future Research Directions}

(1). Ambient Internet of Things (IoT) with Zero-Energy Devices: Compared to the conventional multiple access schemes shown in Figs. 1(a) and 1(c),  NGMA with the three aforementioned  features requires  each user to carry out more transmissions. For energy constrained communication networks,   a device that carries out many  transmissions can suffer a short battery lifespan.  A promising solution for this energy consumption issue  is to apply ambient IoT with zero-energy devices, where backscatter communication (BackCom) can be applied to ensure that the NOMA transmission is carried out in a battery-less manner. This  ambient backscatter feature may be particularly useful  for NGMA, as explained in the following. Take the four-user scenario shown in Figs. 1(e) and 1(f) as an example. Assume that TDMA has been deployed in the legacy network. Hybrid NOMA can be implemented in an ambient manner, where NOMA transmission is powered by  the existing   transmission in the legacy network. For example, user $2$ wants to carry out a NOMA transmission in the first time slot, as shown in Fig. 1(f). To avoid  draining its own battery  by the NOMA transmission, user $2$  can reflect and modulate   the signal sent by user $1$, i.e., user $1$'s signal is treated as a carrier signal by user $2$. Therefore, from the communication perspective,  the use of BackCom can effectively enhance the energy and spectrum cooperation among the users in  NGMA networks; however, from the optimization perspective, the system optimization becomes more challenging, e.g., in addition to the users' transmit powers, the users' reflection  coefficients also need to be optimized, and these system parameters are strongly  coupled.  

(2). Near-Field Communications: In addition to NOMA,   SDMA has also  been envisioned to be a key component of NGMA, since SDMA can efficiently utilize the spatial degrees of freedom offered by   multiple-input multiple-output (MIMO) systems. For conventional far-field MIMO systems, when  the users' distances are larger than the Rayleigh distance,  the combination of NOMA and SDMA is an obvious  solution, as explained in the following. For   far-field users,  the plane-wave channel model can be used, which  means that  a user's far-field channel vector is solely parametrized  by its angle of departure/arrival. This is the reason why    a  far-field beamformer is expected to cover   a cone-shaped area, as illustrated in Fig. 1(c). Any two users that  fall in  the same cone-shaped area will have highly correlated channel vectors, and hence can be  served by a single beamformer via NOMA. However, if the millimeter-wave (mmWave) or terahertz (THz) bands are used, extreme massive MIMO can be implemented, which means that the Rayleigh distance becomes quite large, e.g., with a $513$-element uniform linear array (ULA) and a carrier frequency of $30$ GHz, the Rayleigh distance is around $1400$ meters [5]. For   users which are located in the near-field region, the spherical-wave channel model needs to be used, which results in the so-called beam-focusing phenomenon. In particular, a user's channel vector is parametrized not only by its angle of departure/arrival but also by its distance from the base station. Therefore, two near-field  users' channel vectors might not be highly correlated, which makes the   implementation of  the beam sharing concept  shown  in Fig. 1(c) difficult.   One recent study shows that the resolution of near-field beamforming, which measures the correlation  between the   users' channel vectors, is far from perfect [6]. For example, for users which are moderately close to the base station, e.g., their distances  from the base station are  on the order of  the Rayleigh distance, the users' channel vectors can still be highly correlated, and hence the concept of beam sharing is still applicable to these near-field users. In addition, NOMA assisted NGMA can also play an important role to enable the coexistence of near-field and far-field communications. 

(3). Exploiting Heterogenous  Channel Conditions: Unlike OMA, exploiting the users' heterogenous channel conditions is inherent in the design of NOMA. Take two-user power-domain NOMA as an example, where the user with the weaker channel conditions is allocated more transmit power. Actually,     the users' heterogenous channel conditions are key to the performance gain of power-domain NOMA over OMA, i.e., the performance gap between the two multiple access schemes is zero if the users have the same channel conditions.  However, for NOMA assisted NGMA,   it is challenging to exploit  the users' heterogenous channel conditions, due to its multi-dimensional nature. For example, a typical sum-rate maximization problem for NOMA assisted NGMA is given as follows:
  \begin{problem}\label{pb:3} 
  \begin{alignat}{2}
\underset{   }{\rm{max}} &\quad   \sum^{N}_{n=1}  \sum^{M}_{i=1} R_{i,n}  \quad s.t. &\quad  \ \sum^{N}_{n=1}  \sum^{M}_{i=1}P_{i,n}   \leq   P_{\rm total},  \nonumber
  \end{alignat}
\end{problem}  
\noindent where $R_{i,n}$ denotes user $i$'s data rate in time slot $n$, $P_{i,n}$ denotes the user's corresponding transmit power, $P_{\rm total}$ denotes the overall power budget,  $N$ denotes the total number of time slots, and $M$ denotes the number of users. The challenge in solving  Problem \ref{pb:3} is the availability of the users'   channel state information (CSI), which is the reason why the users' CSI was assumed to be constant   in [4]. This assumption is valid only for  the case of low-mobility users.  A promising solution is to apply  ISAC, where the users' CSI can be estimated and predicted  for carrying out dynamic resource allocation. 

(4). Dynamic Long-Term System Optimization:  As discussed previously, NOMA assisted NGMA needs to be designed by fully utilizing multiple domains. Problem \ref{pb:3} shows a typical example of multi-time-slot opimization.  For the case with a finite number of time slots, this optimization problem can be solved by applying conventional convex optimization tools. However,   for the case with $N\rightarrow \infty$, Problem \ref{pb:3} becomes challenging to solve. We note that such long-term system optimization of	 NGMA is important since NGMA is expected to enable a user to  dynamically  switch between different multiple access modes over time.   One promising solution to achieve dynamic long-term system optimziation is to apply reinforcement learning by modeling the resource allocation as a Markov decision process, i.e., Problem \ref{pb:3} is recast as follows:
  \begin{problem}\label{pb:4} 
  \begin{alignat}{2}
\underset{   }{\rm{max}} &\quad   \mathcal{E}\left\{\left.\sum^{\infty}_{n=1} \gamma ^{n-1} \sum^{M}_{i=1} R_{i,n} \right|  \pi_{P_{i,n}},s_0\right\}    ,\nonumber
  \end{alignat}
\end{problem}  
\noindent where $\gamma$ denotes a discount rate parameter,  $ \pi_{P_{i,n}}$ denotes a policy which is a set of sequential decisions for $P_{i,n}$,  the conditional expectation term $ \mathcal{E}\left\{\cdot \right\}$ denotes the expected value of the discounted cumulative sum of the system throughput if   policy   $\pi_{P_{i,n}}$ is used, and $s_0$ denotes the initial network state, e.g.,  how many bits each user needs to deliver, the users' transmit power budgets, the users' initial CSI,  etc.  We note that this reinforcement learning approach could also be  useful to efficiently exploit the users' dynamic channel conditions without the global CSI  assumption, since reinforcement learning is robust to the changes in the environment. 
 
{\it Conclusions:} NOMA assisted NGMA has been envisioned in the recently published IMT-2030 Framework. This perspective has outlined three important features of NOMA assisted NGMA, namely multi-domain utilization, multi-mode compatibility, and multi-dimensional optimality, where  important directions for future research into the design of NOMA assisted NGMA have also been discussed.

\Acknowledgements{This work was supported by xxxx.}



\end{multicols}
\end{document}